\documentclass[11pt]{article}
\usepackage{amssymb}

\begin{document}

\title{ Electroweak Corrections to the Charged Higgs Boson Decay into Chargino
and Neutralino \footnote{Supported by National
Natural Science Foundation of China.}} \vspace{3mm}

\author{{ Wan Lang-Hui$^{b}$, Ma Wen-Gan$^{a,b}$,Zhang Ren-You$^{b}$, and Jiang Yi$^{b}$ }\\
{\small $^{a}$ CCAST (World Laboratory), P.O.Box 8730, Beijing 100080,P.R.China} \\
{\small $^{b}$ Department of Modern Physics, University of Science and Technology}\\
{\small of China (USTC), Hefei, Anhui 230027, P.R.China} }

\date{}
\maketitle
\vskip 12mm

\begin{abstract}
The electroweak corrections to the partial widths of the $H^+
\rightarrow \tilde{\chi}^+_i \tilde{\chi}_j^0~(i=1,j=1,2)$ decays
including one-loop diagrams of the third generation quarks and
squarks, are investigated within the Supersymmetric Standard
Model. The relative corrections can reach the values about $10\%$,
therefore they should be taken into account for the precise
experimental measurement at future colliders.
\end{abstract}

\vskip 5cm


\vfill \eject

\baselineskip=0.36in

\renewcommand{\thesection}{}
\newcommand{\nb}{\nonumber}
\newcommand{\mch}{m_{\tilde{\chi}^+_i}}
\newcommand{\mno}{m_{\tilde{\chi}^0_j}}

\makeatletter      
\@addtoreset{equation}{section}
\makeatother       

\par
Over the past few years, much efforts have been devoted to search
the new physics beyond the standard model (SM) \cite{s1}\cite{s2}.
The minimal supersymmetric model(MSSM) \cite{s3} is considered as
one of the most attractive extended models. The MSSM predicts
three neutral Higgs bosons, two CP even bosons ($h^0,H^0$) one CP
odd boson ($A^0$), and two charged Higgs bosons ($H^{\pm}$). The
existence of $H^{\pm}$ bosons would provide conclusive evidence of
physics beyond the SM. Therefore, searching for charged Higgs
bosons is one of the major tasks at present and future colliders.
\par
In the future Linear Colliders (LC) there are several mechanisms
which can produce charged Higgs bosons: (1) Pair production via
$e^+e^- \to H^{+}H^{- }$ \cite{eech1} and $e^+e^- \to \gamma
\gamma \to H^{+}H^{-}$\cite{Ma1}. It is found that these two
processes are prominent in discovery of the charged Higgs bosons.
(2) Single charged Higgs boson production in association with
$W^{\pm}$ gauge boson at $e^+e^-$ colliders, provides an
attractive alternative in searching for $H^{\pm}$, which is
kinematically favored when $m_{H^{\pm}}$ exceeds
$m_W$\cite{zf1}.
\par
 At the CERN Large Hadron Collider(LHC),
charged Higgs boson pair can be produced via Drell-Yan process $q
\bar{q} \rightarrow H^{+}H^{-}$ \cite{Desh} and gluon-gluon fusion
$gg \rightarrow H^{+}H^{-}$ at one-loop order\cite{jiang}, when
the charged Higgs boson is not too heavy. For a charged
Higgs boson with mass $m_{H^{\pm}} < m_t-m_b$, the $H^{\pm}$
bosons can abundantly be
produced in decays of top (anti-top) quarks $t(\bar{t}) \rightarrow bH^{+} (
\bar{b}H^{-})$ from the parent production channel $pp\rightarrow
t\bar{t}$. The dominant decay channels in this mass range are
$H^{\pm} \rightarrow \tau^{\pm}\nu_{\tau}$. When $m_{H^{\pm}}>
m_t+m_b$, the dominant production of single heavy charged Higgs
boson is via $gb(g\bar{b}) \rightarrow t H^{-} (\bar{t} H^{+})$
\cite{Gun}, $gg \rightarrow t\bar{b}H^{-} (\bar{t} bH^{+})$
\cite{Diaz} and $qb(\bar{q} \bar{b}) \rightarrow
q^{\prime}bH^{+}(\bar{q}^{\prime}\bar{b}H^{+})$\cite{Oda}
processes. The sequential decay $H^{+} \rightarrow t\bar{b}$ is
known as a preferred channel for $H^{\pm}$ boson search. But this
charged Higgs boson signal appears together with large QCD
background. Alternative production modes of heavy charged Higgs
bosons are associated heavy $H^{\pm}$ production with $W^{\mp}$
bosons via $gg,q \bar{q} \rightarrow W^{\pm} H^{\mp}$. In this
case the $W^{\pm}$-boson's leptonic decay may be used as a
spectacular trigger. This associated heavy $H^{\pm}$ productions
with $W^{\mp}$-boson have been investigated in Refs.\cite{glwh}
\cite{glwh1} \cite{glwh2}.
\par
The heavy charged Higgs boson decay modes in the SM (including
$H^{-}\rightarrow b \bar{t}$,$H^{-}\rightarrow s
\bar{c}$,$H^{-}\rightarrow W^- h$ and $H^{-}\rightarrow \tau
\bar{\nu_{\tau}}$) have be studied and simulated in
Refs.\cite{gunion2}, \cite{Assamagan} and the references therein.
In Ref.\cite{Assamagan} it is concluded that $H^{\pm}$ with mass
up to $\sim 400GeV$ can be discovered by the LHC, if $\tan\beta
\lesssim 3$ or $\tan\beta \gtrsim 25$. The SUSY decay modes
$H^{\pm}\rightarrow \tilde{\chi}^{\pm}_i
\tilde{\chi}^{0}_j~(i=1,~j=1,2,3)$ as the signatures of the MSSM
charged Higgs boson at the LHC have been discussed in
Ref.\cite{Bisset}, there they simulated the decay
$H^{\pm}\rightarrow\tilde{\chi}^{\pm}_i
\tilde{\chi}^{0}_j~(i=1,~j=1,2,3)$ in the regions of the parameter
space, where the identification of $H^{\pm}$ via their SM decay
modes have been proven to be ineffective at the LHC. The MSSM
decay $H^{\pm} \rightarrow \tilde{\chi}^{\pm}_1
\tilde{\chi}^{0}_1$ yields a single hard lepton from the chargino
sequential decay, and $H^{\pm} \rightarrow
\tilde{\chi}^{\pm}_1\tilde{\chi}^{0}_2,~
(\tilde{\chi}^{\pm}_1\tilde{\chi}^{0}_3)$ decays yield three
leptons from chargino and neutralino decays.
\par
As we know that for the comparison of the theoretical prediction
with the precision experiments at the LHC, calculations at
tree-level are no more sufficient, the inclusion of radiative
corrections is necessary. In this paper, we investigate the
one-loop electroweak corrections with top(stop) and
bottom(sbottom) contributions to the heavy charged Higgs boson
decay $H^{\pm}\rightarrow \tilde{\chi}_{i}^+
\tilde{\chi}_{j}^0~(i=1,~j=1,2)$. Since there is no one-loop QCD
corrections in these processes, so the electroweak corrections
with quarks and squarks are most important. The Feynman diagrams
for the $H^{\pm} \rightarrow \tilde{\chi}^{\pm}_i
\tilde{\chi}^{0}_j$ at the tree-level are shown in figure 1(a).
The one-loop Feynman diagrams involving the third generation
quarks and their SUSY partners in the decay of charged Higgs boson
to chargino and neutralino, are drawn in figures 1(c). The
self-energy of charged Higgs boson, chargino self-energy,
neutralino self-energy and $H^{\pm}-W^{\pm}$ mixing self-energy
are shown in figure 1(d-g) respectively.
\par
In the calculation, we use the t' Hooft gauge and adopt the
dimensional reduction scheme(DR)\cite{Copper}, which is commonly
used in the calculation of the electroweak correction in frame of
the MSSM as it preserves supersymmetry at least at one-loop order,
to control the ultraviolet divergences in the virtual loop
corrections. The complete on-mass-shell scheme
\cite{denner}\cite{hollik} is used in doing renormalization.
\par
Firstly, we review briefly the chargino, neutralino, and squark
sectors of the MSSM to settle the conventions. The tree-level mass
matrices for chargino and neutralino can be expressed below
\cite{Haber2}:
\begin{eqnarray}\label{mxn}
X &=& \left( \begin{array}{cc} M & \sqrt{2}m_W \sin\beta\\
                          \sqrt{2}m_W\cos\beta & \mu
\end{array} \right), \nb\\
Y&=& \left(\begin{array}{cccc}
 M^{'} & 0 & -m_Z s_W\cos\beta & m_Z s_W \sin\beta  \\
 0 & M & m_Z c_W\cos\beta & -m_Z c_W \sin\beta \\
-m_Z s_W\cos\beta & m_Z c_W\cos\beta & 0 & -\mu \\
m_Z s_W \sin\beta & -m_Z c_W \sin\beta & -\mu & 0
\end{array} \right).
\end{eqnarray}
Here $s_W\equiv \sin \theta_W, c_W \equiv \cos \theta_W, t_W
\equiv \tan\theta_W$. $M$ and $M^{'}$ are the $SU(2)$ and $U(1)_Y$
soft-SUSY-breaking gaugino masses. These matrices can be
diagonalized by unitary matrices $U,V,N$ via
\begin{eqnarray}
U^{*}XV^{\dag} &=& M_D =
 {\rm diag}\{m_{\tilde{\chi}^+_1}, m_{\tilde{\chi}^+_2}\},
(0<m_{\tilde{\chi}^+_1}<m_{\tilde{\chi}^+_2}), \nb\\
N^{*}YN^{\dag} &=& M_D^0 =
{\rm diag}\{m_{\tilde{\chi}^0_1},m_{\tilde{\chi}^0_2},
m_{\tilde{\chi}^0_3},m_{\tilde{\chi}^0_4}\},
(0<m_{\tilde{\chi}^0_1}<m_{\tilde{\chi}^0_2}<m_{\tilde{\chi}^0_3}<m_{\tilde{\chi}^0_4}).
\end{eqnarray}
The tree-level stop and sbottom squared-mass matrices are:
\begin{eqnarray}
{\cal M}_{\tilde{t}}^2 = \left( \begin{array}{cc}
M_{\tilde{Q}}^2 + m_t^2 + m_Z^2 \cos 2\beta \left(\frac{1}{2}
- \frac{2}{3}s_W^2\right) &  m_t \left(A_t - \mu \cot \beta\right) \\ m_t
\left(A_t - \mu \cot \beta\right) &
M_{\tilde{U}}^2+m_t^2+\frac{2}{3}m_Z^2\cos 2\beta s_W^2  \end{array}\right), \nb\\
{\cal M}_{\tilde{b}}^2 = \left( \begin{array}{cc}  M_{\tilde{Q}}^2
+ m_b^2 - m_Z^2 \cos 2\beta \left(\frac{1}{2} - \frac{1}{3}
s_W^2\right) & m_b\left(A_b - \mu \tan \beta\right) \\ m_b
\left(A_b - \mu \tan \beta\right) &
M_{\tilde{D}}^2+m_b^2-\frac{1}{3}m_Z^2\cos 2\beta s_W^2
\end{array}\right).
\end{eqnarray}
The parameters $M_{\tilde{Q}},M_{\tilde{U}}$ and $M_{\tilde{D}}$
are the soft-SUSY-breaking masses for the third generation SU(2)
squark doublet $\tilde{Q}=\left(\tilde{t}_L,\tilde{b}_L\right)$,
and the singlets $\tilde{U}=\tilde{t}_R$ and
$\tilde{D}=\tilde{b}_R$, respectively. $A_{t,b}$ are the
corresponding soft-SUSY-breaking trilinear couplings. We define
the squark mass eigenstates as following:
\begin{equation}
\left( \begin{array}{cc} \tilde{q}_1 \\ \tilde{q}_2 \end{array}
\right) = R^{\tilde{q}}\left( \begin{array}{cc} \tilde{q}_L \\
\tilde{q}_R \end{array} \right),~~R^{\tilde{q}} = \left(
\begin{array}{cc}\cos\theta_{\tilde{q}} &
 \sin\theta_{\tilde{q}}\\-\sin\theta_{\tilde{q}} & \cos\theta_{\tilde{q}} \end{array}
 \right),~~~q=(t,b),
\end{equation}
$R^{\tilde{q}}$ is used to diagonalize the squark mass matrix
$R^{\tilde{q}}{\cal M}_{\tilde{q}}^2 R^{\tilde{q}\dag} =
\rm{diag}\{m_{\tilde{q}_1}^2,m_{\tilde{q}_2}^2\}$.
\par
The vertexes of $H^+\tilde{\chi}^+ \tilde{\chi}^0$ and
$G^+\tilde{\chi}^+ \tilde{\chi}^0$, which are related to
$H^+\rightarrow\tilde{\chi}^+_i \tilde{\chi}^0_j ~(i=1,~j=1,2)$
partial decay width are determined by the interaction Lagrangian,
which are given by\cite{Haber2}:
\begin{eqnarray} \label{lg}
{\cal L}_{H^+\tilde{\chi}^+ \tilde{\chi}^0} = - H^+
\bar{\tilde{\chi}}_i^+ \left[g \sin \beta Q^{'R*}_{ij}P_L + g \cos
\beta Q^{'L}_{ij} P_R \right] \tilde{\chi}_j^0 + h.c.,
\end{eqnarray}
\begin{eqnarray} \label{lh}
{\cal L}_{G^+\tilde{\chi}^+ \tilde{\chi}^0} = -G^+
\bar{\tilde{\chi}}_i^+ \left[-g \cos \beta Q^{'R*}_{ij}P_L + g
\sin \beta Q^{'L}_{ij} P_R \right] \tilde{\chi}_j^0 + h.c. ,
\end{eqnarray}
where$P_{L,R} = \frac{1}{2}(1\mp\gamma_5)$ are the chirality
projection operators, and we denoted:
\begin{eqnarray}
Q^{'R}_{ij} = N_{j3}U_{i1} - \sqrt{\frac{1}{2}}(N_{j2}+N_{j1}\tan \theta_W)U_{i2}, \nb \\
Q^{'L}_{ij} = N_{j4}V_{i1} + \sqrt{\frac{1}{2}}(N_{j2}+N_{j1}\tan
\theta_W)V_{i2}.
\end{eqnarray}
The decay width of $H^+ \rightarrow \tilde{\chi}^+_i
\tilde{\chi}^0_j$ at the tree-level is expressed as\cite{Gunion3}
\begin{eqnarray}
\Gamma_{\rm{tree}}&=&\frac{g^2 \lambda^{1/2}}{16 \pi
m_{H^+}^3}\left[\left(|F^L_{ij}|^2+|F^R_{ij}|^2\right)\left(m_{H^+}^2
- m_{\tilde{\chi}^{+}_i}^2-m_{\tilde{\chi}^{0}_i}^2
\right)-4m_{\tilde{\chi}^+_i}m_{\tilde{\chi}^0_j}Re\left(F^L_{ij}F^{R}_{ij} \right)\right], \nb \\
\end{eqnarray}
where
\begin{eqnarray}
\lambda^{1/2}&=&\left[\left(m_{H^+}^2 -
m_{\tilde{\chi}^+_i}^2-m_{\tilde{\chi}^0_j}^2 \right)^2 - 4
m_{\tilde{\chi}^+_i}^2 m_{\tilde{\chi}^0_j}^2\right]^{1/2},\nb\\
F^R_{ij} &=& \sin \beta Q^{'R}_{ij},~~~~~~ F^L_{ij} = \cos \beta
Q^{'L}_{ij}.
\end{eqnarray}
To obtain the renormalized vertex
$H^+\tilde{\chi}^+_i\tilde{\chi}^0_j$, we introduce parameters and
field renormalization constants as follow\cite{denner,hollik}:
\begin{eqnarray}\label{rc}
g &\rightarrow& \left(1+\frac{\delta g}{g}\right)g = \left( 1+ \delta Z_e -
 \frac{\delta s_W}{s_W} \right) g, \nb \\
\tan \beta &\rightarrow& \left( 1+ \frac{\delta \tan \beta}{\tan \beta}\right)\tan \beta,
~~t_W \rightarrow \left(1 +\frac{\delta t_W}{t_W}\right) t_W, \nb \\
\left(
\begin{array}{cc}
 H^\pm \\ G \pm
\end{array}
\right) &\rightarrow& \left(
\begin{array}{cc}
1+\frac{1}{2}\delta Z_{H^\pm} & \frac{1}{2} \delta Z_{H^\pm G^ \pm} \\
\frac{1}{2} \delta Z_{G^\pm H^\pm} & 1 + \frac{1}{2}\delta Z_{G^\pm}
\end{array}
\right) \left( \begin{array}{cc} H^\pm \\ G^\pm \end{array} \right) \nb\\
 \tilde{\chi}^+_i &\rightarrow&
\left(\delta_{ik} + \frac{1}{2} \delta Z_{+,ik}^LP_L + \frac{1}{2} \delta
Z_{+,ik}^R P_R \right)\tilde{\chi}^+_k, \nb\\
\tilde{\chi}^0_i &\rightarrow& \left(\delta_{ik} + \frac{1}{2} \delta Z_{0,ik}^LP_L + \frac{1}{2} \delta
Z_{0,ik}^R P_R \right)\tilde{\chi}^0_k, \nb\\
U &\rightarrow& U + \delta U, ~~V \rightarrow V + \delta V,~~ N
\rightarrow N + \delta N,
\end{eqnarray}
here for the charged Higgs boson, chargino and neutralino, we
introduced the mixed field renormalization constants and the
counterterms for the $U,~V$ and $N$ matrices.
\par
In the on-mass-shell scheme the renormalization constants defined
in Eq.(\ref{rc}) can be fixed by the following renormalization
conditions\cite{denner}\cite{hollik}:
\par
\begin{itemize}
\item  The renormalized tadpoles, i.e.~the sum of
      tadpole diagrams $T_{h,H}$ and tadpole counter-terms
      $\delta_{h,H}$ vanish:
\[   T_{h} +\delta t_h=0, \quad T_H +\delta t_H=0 \, . \]
So no tadpole needed to be considered in our calculation.
\item On shell condition for the charged Higgs boson,
\begin{eqnarray}
\tilde{Re}\hat{\Sigma}_{H^{\pm}H^{\pm}}(m_{H^+}^2)=0,
~~\tilde{Re}\hat{\Sigma}_{H^{\pm}H^{\pm}}^{'}(m_{H^+}^2)=0,
\end{eqnarray}
the "hat" denotes the renormalized quantities,
$\Sigma^{'}(m^2)\equiv\frac{\partial \Sigma(p^2)}{\partial
p^2}|_{p^2 = m^2}$. $\tilde{Re}$ takes only the real part of the
loop integrals appearing in the self-energies. From which we
abtain
\begin{eqnarray}
\delta m_{H^{\pm}}^2 = \tilde{Re} \Sigma_{H^{\pm}H^{\pm}}(m_{H^+}^2),~~\delta Z_{H^{\pm}} =
-\tilde{Re} \Sigma_{H^{\pm}H^{\pm}}^{'}(m_{H^+}^2).
\end{eqnarray}
\item The fermion one-loop renormalized two-point function can be decomposed as
\begin{equation}
i \hat{\Gamma}^f_{ij}(p) = i \delta_{ij}(\rlap/{p} - m_{f_i}) +
i \left[ \rlap/{p}P_L \hat{\Sigma}^L_{ij}(p^2)
+\rlap/{p}P_R \hat{\Sigma}^R_{ij}(p^2)+
 P_L \hat{\Sigma}^{SL}_{ij}(p^2)
 +P_R\hat{\Sigma}^{SR}_{ij}(p^2)\right].
\end{equation}
$\hat{\Sigma}^{L,R,SL,SR}_{ij}$ are the fermions self-energy
matrices. By imposing the on-shell renormalization conditions for
fermions, We can obtain the renormalization constants as
following:
\begin{eqnarray}\label{ff}
\delta m_i &=& \frac{1}{2}\tilde{Re}\left[m_i(\Sigma^L_{ii}(m_i^2)+\Sigma^R_{ii}(m_i^2))+
\Sigma^{SL}_{ii}(m_i^2)+ \Sigma^{SR}_{ii}(m_i^2)\right], \nb \\
\delta Z^L_{ii}&=&-\tilde{Re}\left[\Sigma^L_{ii}(m_i^2) + m_i^2(\Sigma^{L'}_{ii}(m_i^2)+\Sigma^{R'}_{ii}(m_i^2)) +
                         m_i (\Sigma^{SL'}_{ii}(m_i^2) + \Sigma^{SR'}_{ii}(m_i^2))\right], \nb \\
\delta Z^R_{ii}&=&-\tilde{Re}\left[\Sigma^R_{ii}(m_i^2) + m_i^2(\Sigma^{L'}_{ii}(m_i^2)+\Sigma^{R'}_{ii}(m_i^2)) +
                         m_i (\Sigma^{SL'}_{ii}(m_i^2) + \Sigma^{SR'}_{ii}(m_i^2))\right], \nb \\
\delta Z^L_{ij}&=&\frac{2}{m_i^2 - m_j^2}\tilde{Re}\left[m_j^2\Sigma^L_{ij}(m_j^2)+ m_i m_j\Sigma^{R}_{ij}(m_j^2)+
                          m_i \Sigma^{SL}_{ij}(m_j^2) + m_j \Sigma^{SR}_{ij}(m_j^2) \right], ~~~\rm{for}~ i \neq j\nb \\
\delta Z^R_{ij}&=&\frac{2}{m_i^2 - m_j^2}\tilde{Re}\left[m_i m_j\Sigma^L_{ij}(m_j^2)+ m_j^2\Sigma^{R}_{ij}(m_j^2)+
                          m_j \Sigma^{SL}_{ij}(m_j^2) + m_i \Sigma^{SR}_{ij}(m_j^2) \right]. ~~~\rm{for}~ i \neq j\nb \\
\end{eqnarray}
Then we can obtain the chargino wave function renormalization
constants $\delta Z_{+,ij}$ and neutralino wave function
renormalization constants $\delta Z_{0,ij}$ from above equations.

\item The renormalization constant for the electric charge and counterterms
of gauge boson masses:
\begin{eqnarray}
\delta Z_e &=& -\frac{1}{2}\delta Z_{\gamma\gamma} - \frac{s_W}{c_W}\frac{1}{2}\delta
Z_{Z\gamma}=\frac{1}{2}\Sigma_{\gamma\gamma}^{T'}(0) - \frac{s_W}{c_W}
\frac{\Sigma_{\gamma Z}^T(0)}{m_Z^2}, \nb \\
\delta m_{V}^2 &=& \tilde{Re} \Sigma_{VV}^T(m_V^2),~~~(V = W^{\pm},Z).
\end{eqnarray}
For the weak mixing angle we use the definition $\cos^2 \theta_W =
\frac{m_W^2}{m_Z^2}$, this gives:
\begin{equation}
\frac{\delta c_W}{c_W} =\frac{1}{2}\left(\frac{\delta m_W^2}{m_W^2} - \frac{\delta m_Z^2}{m_Z^2}\right).
\end{equation}
\item  The tree level Lagrangian contains a mixing
of the gauge boson fields $W^{\pm}$ and the Goldstone boson fields
$G^{\pm}$, namely
\begin{eqnarray}
{\cal L}_{GW} = im_W W_{\mu}^- \partial^{\mu} G^{+} + h.c..
\end{eqnarray}
After substituting the renormalization  transformation for the
Goldstone defined in Eq.(\ref{rc}), We can obtain the relevant
counterterms
\begin{eqnarray}
\delta {\cal L}_{HW} = \frac{1}{2}\delta Z_{G^{+}H^{+}} i
m_W W_{\mu}^{-} \partial^{\mu} H^{+} + h.c..
\end{eqnarray}
Then the renormalized one-particle-irreducible two-point Green
function can be written as:
\begin{eqnarray}
i \hat{\Gamma}^{\mu}_{HW}(k) = i\left(\Sigma_{H^{+}W^{+}}(k^2) +
\frac{1}{2}m_W \delta Z_{G^{+}H^{+}}\right) k^{\mu} .
\end{eqnarray}
We impose the renormalization condition in a way that the physical
charged Higgs boson dose not mix with the physical $W$ boson, then
we have
\begin{eqnarray}
\tilde{Re}\hat{\Gamma}_{HW}(k)|_{k^2=m_{H^{+}}^2} = 0
\Longrightarrow \delta Z_{G^{+}H^{+}} = \frac{-2 \tilde{Re} \Sigma_{H^+W^+}(m_{H^+}^2)}{m_W}.
\end{eqnarray}
 If we impose $\delta v_1/v_1 = \delta v_2/v_2$, analogous to the Eq.(A.15) of
Ref.\cite{Eberl}, We have:
\begin{equation}
\frac{\delta \tan \beta}{\tan \beta} = -\frac{1}{m_W \sin 2
\beta}\tilde{Re}\Sigma_{H^{+}W^{+}}(m_{H^{+}}^2),
\end{equation}
then the counterterm $\delta\tan\beta$ can also be obtain from the
$H^{+}-W^{+}$ mixing self-energy.
\item The counterterms $\delta U, \delta V$ and $\delta N$
can be fixed by requiring that the counterterms
$\delta U, \delta V$, and $\delta N$ cancel the antisymmetric
parts of the wave function corrections
\cite{Eberl}\cite{Denner2}\cite{Kniehl}:
\begin{eqnarray}
\delta U = \frac{1}{4}\left( \delta Z_{+}^R - \delta Z_{+}^{R \dag} \right)U, \nb \\
\delta V = \frac{1}{4}\left( \delta Z_{+}^L - \delta Z_{+}^{L \dag} \right) V, \nb \\
\delta N = \frac{1}{4}\left( \delta Z_{0}^L - \delta Z_{0}^{L \dag} \right) N.
\end{eqnarray}
\par
The explicit expressions of all the self-energies concerned in our
calculation, can be found in the Appendix and the cited references
therein.
\end{itemize}
\par
By substituting Eq.(\ref{rc}) into the bare Lagrangian
Eq.(\ref{lg}), we can obtain the counterterms as follow:
\begin{eqnarray}\label{ct}
\delta {\cal L}_{H^+\tilde{\chi}^+ \tilde{\chi}^0} &\equiv&
 -H^+ \bar{\tilde{\chi}}_i^+ \left[\delta C^L_{ij} P_L +
\delta C^R_{ij} P_R \right] \tilde{\chi}_j^0 \nb\\
&=&-H^+ \bar{\tilde{\chi}}_i^+ \left\{
  g \sin \beta \left[ Q^{'R*}_{ij}\left(\delta Z_e - \frac{\delta s_W}{s_W}
+ \frac{\delta \sin \beta}{\sin\beta} + \frac{1}{2} \delta Z_{H^+}
- \frac{1}{2}\cot \beta \delta Z_{G^+H^+} \right) \right. \right. \nb \\
 &+& \left.  \sum_{k=1}^{2}\frac{1}{2}Q^{'R*}_{kj}\delta Z^R_{+,ki}
+ \sum_{k=1}^{4}\frac{1}{2}Q^{'R*}_{ik}\delta Z^L_{0,kj}
+ \delta Q^{'R*}_{ij} \right]P_L \nb \\
&+& g \cos \beta \left[Q^{'L}_{ij} \left(
\delta Z_e - \frac{\delta s_W}{s_W} + \frac{\delta \cos \beta}{\cos\beta}
+ \frac{1}{2} \delta Z_{H^+} + \frac{1}{2}\tan \beta \delta Z_{G^+H^+} \right) \right. \nb \\
&+& \left. \left. \sum_{k=1}^{2}\frac{1}{2}Q^{'L}_{kj}\delta Z^L_{+,ki}
+ \sum_{k=1}^{4}\frac{1}{2}Q^{'L}_{i,k}\delta Z^R_{0,kj}
+ \delta Q^{'L}_{ij} \right] P_R \right\} \tilde{\chi}^0_j,
\end{eqnarray}
where we have defined
\begin{eqnarray}
\delta Q^{'R}_{ij} &=& \delta N_{j3}U_{i1} + N_{j3} \delta U_{i,1}
-\frac{1}{\sqrt{2}}\left(\delta N_{j2} + \delta
N_{j1} t_W + N_{j1} \delta t_W\right) U_{i,2} -\frac{1}{\sqrt{2}}\left( N_{j2}+N_{j1} t_W \right) \delta U_{i2}, \nb \\
\delta Q^{'L}_{ij} &=& \delta N_{j4}V_{i1} + N_{j4} \delta V_{i,1}
+\frac{1}{\sqrt{2}}\left(\delta N_{j2} + \delta N_{j1} t_W +
N_{j1} \delta t_W\right) V_{i,2}
+\frac{1}{\sqrt{2}}\left(N_{j2}+N_{j1}t_W \right) \delta V_{i2}.
\nb\\
\end{eqnarray}
Then the renormalized one-loop part of the amplitude for the decay
$H^+ \rightarrow \tilde{\chi}^+_i \tilde{\chi}^0_j$ can be written
as
\begin{equation}\label{rm}
{\cal M}^{Loop}_{ij} = -i H^+ \bar{\tilde{\chi}}_i^+
\left[ C^L_{ij} P_L +  C^R_{ij} P_R \right]
\tilde{\chi}_j^0;~~~C^{L,R}_{ij}=\delta C^{L,R}_{ij}+\Lambda^{L,R}_{ij}.
\end{equation}
where $\delta C^{L,R}_{ij}$ are defined in Eq.(\ref{ct}), and
$\Lambda^{L,R}$ are the form factors contributed by the diagrams
in Fig.1.(c), their expressions are listed in the Appendix. We
have checked both analytically and numerically that the
renormalized amplitude Eq.(\ref{rm}) is ultraviolet finite.
\par
We define the relative correction quantitatively as a ratio of the
decay width correction from one-loop diagrams to the tree-level
width:
\begin{equation}
\delta^{ij} =\frac{\Gamma_{\rm total} (H^+\rightarrow
\tilde{\chi}^+_i \tilde{\chi}^0_j) -
\Gamma_{\rm{tree}}(H^+\rightarrow \tilde{\chi}^+_i
\tilde{\chi}^0_j)}{\Gamma_{\rm{tree}}(H^+\rightarrow
\tilde{\chi}^+_i \tilde{\chi}^0_j)}.
\end{equation}
\par
Now we turn to the numerical analysis. The SM input parameters are
chosen as: $m_t=174.3~GeV$,$m_b=4.3~GeV$, $m_{Z}=91.1882~GeV$,
$m_{W}=80.419~GeV$ and $\alpha_{EW} = 1/128$\cite{pdg}. The SUSY
parameters are taken as follows by default unless otherwise
stated. We choose
$M_{\tilde{Q}}=M_{\tilde{U}}=M_{\tilde{D}}=A_t=A_b= 200 GeV$. The
ratio of the vacuum expectation values $\tan\beta$ is set to be 4
or 30 in order to make comparison. The mass of charged Higgs boson
$m_{H^{\pm}} = 250~GeV$. The physical chargino masses
$m_{\tilde{\chi}_1^+}$, $m_{\tilde{\chi}_2^+}$ and the lightest
neutralino mass $m_{\tilde{\chi}_1^0}$ are set to be $100~GeV$,
$300~GeV$ and $60~GeV$, respectively. Then the fundamental SUSY
parameters $M$, $M^{'}$ and $\mu$ in Eq.(\ref{mxn}) can be
extracted from these input chargino masses, lightest neutralino
mass $m_{\tilde{\chi}_1^0}$ and $\tan\beta$. Here we assume $\mu$
has negative sign.
\par
In Fig.2 we plot the relative corrections as functions of
$\tan\beta$ for $H^+\rightarrow\tilde{\chi}^+_1\tilde{\chi}^0_1$
and $H^+ \rightarrow \tilde{\chi}^+_1\tilde{\chi}^0_2$. We can see
that the relative electroweak corrections for $H^+ \rightarrow
\tilde{\chi}^+_1\tilde{\chi}^0_1$ are always positive, while
corrections for $H^+\rightarrow \tilde{\chi}^+_1\tilde{\chi}^0_2$
decay are negative. The corrections increase as the increment of
$\tan\beta$ due to the coupling enhancement at large $\tan\beta$.
Quantitatively the relative corrections can reach about $10\%$ and
$-8\%$ for $\delta^{11}$ and $\delta^{12}$ respectively when
$\tan\beta$ is about 36.
\par
In Fig.3 we present the plot of the relative corrections versus
the mass of charged Higgs boson. We can see there are some peaks
in the figure, which corresponding to the resonance effects at the
positions where $m_{H^{\pm}}(346.2~GeV) =
m_{\tilde{t}_1}(181.1~GeV)+m_{\tilde{b}_1} (165.1~GeV)$,
$m_{H^{\pm}}(419.6GeV)= m_{\tilde{t}_1}(181.1~GeV)
+m_{\tilde{b}_2}(238.5~GeV)$, $m_{H^{\pm}}(487.2~GeV)=
m_{\tilde{t}_2}(322.1~GeV)+m_{\tilde{b}_1}(165.1~GeV)$ for
$\tan\beta=30$, and  $m_{H^{\pm}}(369.2~GeV) =
m_{\tilde{t}_1}(171.7~GeV)+m_{\tilde{b}_1} (197.5~GeV)$,
$m_{H^{\pm}}(383.1~GeV)= m_{\tilde{t}_1}(171.7~GeV)
+m_{\tilde{b}_2}(211.4~GeV)$ for $\tan\beta=4$, respectively. For
fixed value of $\tan\beta$, the relative correction is insensitive
to the variation of the charged Higgs boson mass in the region
between $250~GeV$ and $450~GeV$ except around the positions at
resonance masses. But there is a sharp peak at the position
$m_{H^{\pm}}=487.2~GeV$ due to the resonance effect, and the
correction can vary from $8\%$ to $-3\%$.
\par
The relative corrections of
$H^+\rightarrow\tilde{\chi}^+_1\tilde{\chi}^0_1$ versus the
lightest chargino mass with $\tan\beta=4$ and $\tan\beta=30$ are
depicted in figure 4. We can see when $\tan\beta=4$, the relative
correction is about $1\%$ for $m_{\tilde{\chi}^+_1}$ varying from
$100~GeV$ to $180~GeV$, while when $\tan\beta=30$, the relative
correction decreases from $6.5\%$ to $3.8\%$ with
$m_{\tilde{\chi}^+_1}$ increasing in the same range. In Fig.5 we
present the correction
$H^+\rightarrow\tilde{\chi}^+_1\tilde{\chi}^0_1$ as a function of
the lightest neutralino mass. We can see that the relative
corrections is insensitive to the variation of
$m_{\tilde{\chi}^0_1}$ for small $\tan\beta$, but for large
$\tan\beta$, the correction varies from $3.5\%$ to $7.3\%$ as the
lightest neutralino mass running from $50~GeV$ to $90~GeV$.
\par
The results in Fig.6, are worked out in the mSUGRA scenario. Of
the five mSUGRA input parameters($m_0$, $m_{1/2}$, $A_0$,
$\tan{\beta}$ and sign of $\mu$), we take $m_0 =200~GeV$,
$m_{1/2}=150~GeV$, $A_0=300~GeV$, the sign of $\mu$ is set to be
negative, and $\tan\beta$ is running from 2 to 35. In our
numerical evaluation to get the low energy scenario from the
mSUGRA, the renormalization group equations (RGE's)\cite{RGE} are
run from the weak scale $m_Z$ up to the GUT scale, taking all
thresholds into account. We use two-loop RGE's only for the gauge
couplings and the one-loop RGE's for the other supersymmetric
parameters. The GUT scale boundary conditions are imposed and the
RGE's are run back to $m_Z$, again taking threshold into account.
In Fig.6 there is a peak at the position $\tan\beta=3$ on the
curve for $\delta^{12}$, which comes from the resonance effect
where $m_{\tilde{t}_1} = m_t + m_{\tilde{\chi}^0_2}$ is satisfied.
We can see that the correction is sensitive to $\tan\beta$ in
mSUGRA scenario, especially when $\tan\beta$ is located in small
or large value regions. The correction is varying from $1.5\%$ to
$-3\%$ for $H^+\rightarrow\tilde{\chi}^+_1\tilde{\chi}^0_1$ and
$4\%$ to $-2\%$ for
$H^+\rightarrow\tilde{\chi}^+_1\tilde{\chi}^0_2$ when $\tan\beta$
goes from 2 to 35.
\par
In Summary, we have computed the electroweak corrections to the
partial widths of the $H^+ \rightarrow \tilde{\chi_i}^+
\tilde{\chi}_j^0~(i=1,j=1,2)$ decays including the third
generation quark and squark one-loop diagrams within the Minimal
Supersymmetric Standard Model. We find that the relative
corrections can be sizeable and reach the order of $10\%$ ,
therefore they should be taken into account in the precise
experiment analysis.

\noindent{\large\bf Acknowledgement:} This work was supported in
part by the National Natural Science Foundation of China(project
number: 19875049, 10005009), the Education Ministry of China and
the Ministry of Science and Technology of China.

\section{Appendix A}
\renewcommand{\theequation}{A.\arabic{equation}}

In this appendix, we list the self-energies of charged Higgs
boson, the mixing self-energy of $H^{\pm}-W^{\pm}$ and form
factors $\Lambda^{L,R}$ contributed by the one-loop Feynman
diagrams (Fig.1(c)). For the self-energies of chargino,
neutralino, $W$ gauge boson and $Z$ gauge boson with quarks and
squarks in loops, one may refer to Ref.\cite{Eberl}, the
self-energies of $\gamma$ and its mixing with $Z$ gauge boson with
quarks and squarks in loop can be found in \cite{zml}.
\par
The explicit expressions of Feynman rules we used in our
calculation can be found in Ref.\cite{Haber2}. We denote the
couplings that chargino, neutralino coupling with quark and squark
in the forms as bellow(The notations are the same as those in
Ref.\cite{zml}):
\begin{eqnarray}
\bar{b}-\tilde{t}_{i}-\bar{\tilde{\chi}}^+_j &:& \left(
V^{(1)}_{b\tilde{t}_i\tilde{\chi}^+_j}P_L +
 V^{(2)}_{b\tilde{t}_i\tilde{\chi}^+_j}P_R\right)C,~~~
\bar{t}-\tilde{b}_{i}-\tilde{\chi}^+_j:V^{(1)}_{t\tilde{b}_i\tilde{\chi}^+_j}P_L
+
V^{(2)}_{t\tilde{b}_i\tilde{\chi}^+_j}P_R, \nb\\
\bar{b}-\tilde{b}_{i}-\tilde{\chi}^0_j&:&V^{(1)}_{b\tilde{b}_i\tilde{\chi}^0_j}P_L
+ V^{(2)}_{b\tilde{b}_i\tilde{\chi}^0_j}P_R,~~~
\bar{t}-\tilde{t}_{i}-\tilde{\chi}^0_j:V^{(1)}_{t\tilde{t}_i\tilde{\chi}^0_j}P_L
+ V^{(2)}_{t\tilde{t}_i\tilde{\chi}^0_j}P_R,
\end{eqnarray}
where $C$ is the charge conjugation operator. The couplings
between $H^{+}$ and quark(squark), and the vertex of $W^{\pm}$
gauge boson coupling with squarks are denoted as:
\begin{eqnarray}
H^+-\bar{t}-b &:&~ V^{(1)}_{H^+tb}P_L +
V^{(2)}_{H^+tb}P_R,~~~~~H^+-\tilde{t}_i-\tilde{b}_j:~
V_{H^+\tilde{t}_i\tilde{b}_j}, \nb\\
H^+-H^--\tilde{t}_i-\tilde{t}_j&:&~V_{H^+H^-\tilde{t}_i\tilde{t}_j},~~~~~~~~~~
H^+-H^--\tilde{b}_i-\tilde{b}_j:~V_{H^+H^-\tilde{b}_i\tilde{b}_j}, \nb\\
W^--\tilde{t}_i(p_1)-\tilde{b}_j(p_2)&:&~
V_{W\tilde{t}_i\tilde{b}_j} (p_1+p_2)^{\mu}.
\end{eqnarray}
The charged Higgs boson self-energy reads:
\begin{eqnarray}
\Sigma_{H^{\pm}H^{\pm}}&=&\frac{-N_c}{8\pi^2}\left[
\left(V^{(1)}_{H^+tb}V^{(1)*}_{H^+tb}+V^{(2)}_{H^+tb}V^{(2)*}_{H^+tb}\right)
\left(-A_0(m_t) + m_b^2 B_0(p,m_b,m_t))+p^2B_1(p,m_b,m_t)\right)\right. \nb\\
 &+&  \left( V^{(1)}_{H^+tb}V^{(2)*}_{H^+tb}+V^{(1)*}_{H^+tb}V^{(2)}_{H^+tb}\right)
  m_b m_t B_0(p,m_b,m_t)
 -  \frac{1}{2}\sum_{\alpha,\beta=1}^{2}|V_{H^+\tilde{t}_{\alpha}\tilde{b}_{\beta}}|^2
 B_0(p,m_{\tilde{b}_{\beta}},m_{\tilde{t}_{\alpha}}) \nb\\
 &+& \left.\frac{i}{2}\sum_{\alpha=1}^{2} \left[
 V_{H^+H^-\tilde{b}_{\alpha}\tilde{b}_{\alpha}} A_0(m_{\tilde{b}_{\alpha}})
+V_{H^+H^-\tilde{t}_{\alpha}\tilde{t}_{\alpha}} A_0(m_{\tilde{t}_{\alpha}})\right]
  \right].
\end{eqnarray}
The $H^{\pm}-W^{\pm}$ mixed self-energy contributed by the
diagrams in Fig.1(g) is expressed as,
\begin{eqnarray}
\Sigma_{H^{\pm}W^{\pm}} &=&\frac{i g
N_c}{8\sqrt{2}\pi^2}\left[m_bV_{H^{+}ud}^{(2)}(B_0(p, m_b, m_t)+B_1(p,m_b, m_t)) +
 m_tV_{H^+ud}^{(1)}B_1(p, m_b, m_t)\right]\nb\\
 &+& \frac{N_c}{16 \pi^2}\sum_{\alpha,\beta=1}^{2}V_{H^+\tilde{t}_{\alpha}\tilde{b}_{\beta}}
   V_{W \tilde{t}_{\alpha}\tilde{b}_{\beta}}[B_0+2B_1](p, \tilde{t}_{\alpha}, \tilde{b}_{\beta}).
\end{eqnarray}
We decompose the form factors $\Lambda^{L,R}$ contributed by the
one-loop Feynman diagrams in Fig.1(c) according to
\begin{eqnarray}
\Lambda^{L,R}=\Lambda^{L,R}_{(1)}+\Lambda^{L,R}_{(2)}+
\Lambda^{L,R}_{(3)}+\Lambda^{L,R}_{(4)},
\end{eqnarray}
where $\Lambda^{L,R}_{(i)}~(i=1,2,3,4)$ are the form
factors contributed by the diagrams in Fig.1(c1), Fig.1(c2),
Fig.1(c3) and Fig.1(c4), respectively.
\begin{itemize}
\item For diagram Fig.1(c.1), we introduce the following notations:
\begin{eqnarray}
F_a^{(1)}=
V_{H^+tb}^{(2)}V_{t\tilde{b}_{\alpha}\tilde{\chi}^+_i}^{(2)*}V_{b\tilde{b}_{\alpha}\tilde{\chi}^0_j}^{(1)},~~~
F_b^{(1)}=
V_{H^+tb}^{(1)}V_{t\tilde{b}_{\alpha}\tilde{\chi}^+_i}^{(1)*}V_{b\tilde{b}_{\alpha}\tilde{\chi}^0_j}^{(2)},\nb\\
F_c^{(1)} =
V_{H^+tb}^{(2)}V_{t\tilde{b}_{\alpha}\tilde{\chi}^+_i}^{(2)*}V_{b\tilde{b}_{\alpha}\tilde{\chi}^0_j}^{(2)},~~~
F_d^{(1)}=
V_{H^+tb}^{(1)}V_{t\tilde{b}_{\alpha}\tilde{\chi}^+_i}^{(1)*}V_{b\tilde{b}_{\alpha}\tilde{\chi}^0_j}^{(1)},\nb\\
F_e^{(1)} =
V_{H^+tb}^{(2)}V_{t\tilde{b}_{\alpha}\tilde{\chi}^+_i}^{(1)*}V_{b\tilde{b}_{\alpha}\tilde{\chi}^0_j}^{(1)},~~~
F_f^{(1)}=
V_{H^+tb}^{(1)}V_{t\tilde{b}_{\alpha}\tilde{\chi}^+_i}^{(2)*}V_{b\tilde{b}_{\alpha}\tilde{\chi}^0_j}^{(2)},\nb\\
F_g^{(1)} =
V_{H^+tb}^{(2)}V_{t\tilde{b}_{\alpha}\tilde{\chi}^+_i}^{(1)*}V_{b\tilde{b}_{\alpha}\tilde{\chi}^0_j}^{(2)},~~~
F_h^{(1)}=
V_{H^+tb}^{(1)}V_{t\tilde{b}_{\alpha}\tilde{\chi}^+_i}^{(2)*}V_{b\tilde{b}_{\alpha}\tilde{\chi}^0_j}^{(1)}.
\end{eqnarray}
The form factors $\Lambda^{L/R}_{(1)}$ contributed by diagram
Fig.1(c1) are:
\begin{eqnarray}\label{c1}
\Lambda^{L}_{(1)}&=&\frac{-iN_c}{16\pi^2}\sum_{\alpha=1}^{2}\left[F_a^{(1)} B_0
- \left(F_d^{(1)} m_b \mch + F_e^{(1)} m_t \mch
+ F_a^{(1)} \mch^2 + F_b^{(1)} \mch \mno \right) C_{11} \right.\nb\\
&-&\left(F_h^{(1)}m_b m_t + F_d^{(1)} m_b \mch + F_f^{(1)} m_t \mno
+  F_b^{(1)} \mch \mno + F_a^{(1)} m_{\tilde{b}_{\alpha}}^2\right) C_0  \nb\\
&-& \left. \left(F_a^{(1)} (\mno^2 - \mch^2)+ F_c^{(1)} m_b \mno
+ F_f^{(1)} m_t \mno - F_d^{(1)} m_b \mch -
F_e^{(1)}m_t \mch \right)C_{12}  \right], \nb\\
\Lambda^{R}_{(1)}&=&\Lambda^{L}_{(1)}(F_a^{(1)}\leftrightarrow
F_b^{(1)},F_c^{(1)}\leftrightarrow
F_d^{(1)},F_e^{(1)}\leftrightarrow F_f^{(1)},
F_g^{(1)}\leftrightarrow F_h^{(1)} ),
\end{eqnarray}
with $B_0 = B_0(p, m_t,m_b)$, $C_{0,11,12}=C_{0,11,12}(k_1, -p_1,
m_{\tilde{b}_{\alpha}}, m_t, m_b).$
\item For diagram Fig.1(c.2), we define the notations as:
\begin{eqnarray}
F_a^{(2)}=
V_{H^+tb}^{(2)}V_{t\tilde{t}_{\alpha}\tilde{\chi}^0_j}^{(2)*}V_{b\tilde{t}_{\alpha}\tilde{\chi}^+_i}^{(1)},~~~
F_b^{(2)}=
V_{H^+tb}^{(1)}V_{t\tilde{t}_{\alpha}\tilde{\chi}^0_j}^{(1)*}V_{b\tilde{t}_{\alpha}\tilde{\chi}^+_i}^{(2)},\nb\\
F_c^{(2)} =
V_{H^+tb}^{(2)}V_{t\tilde{t}_{\alpha}\tilde{\chi}^0_j}^{(1)*}V_{b\tilde{t}_{\alpha}\tilde{\chi}^+_i}^{(1)},~~~
F_d^{(2)}=
V_{H^+tb}^{(1)}V_{t\tilde{t}_{\alpha}\tilde{\chi}^0_j}^{(2)*}V_{b\tilde{t}_{\alpha}\tilde{\chi}^+_i}^{(2)},\nb\\
F_e^{(2)} =
V_{H^+tb}^{(2)}V_{t\tilde{t}_{\alpha}\tilde{\chi}^0_j}^{(2)*}V_{b\tilde{t}_{\alpha}\tilde{\chi}^+_i}^{(2)},~~~
F_f^{(2)}=
V_{H^+tb}^{(1)}V_{t\tilde{t}_{\alpha}\tilde{\chi}^0_j}^{(1)*}V_{b\tilde{t}_{\alpha}\tilde{\chi}^+_i}^{(1)},\nb\\
F_g^{(2)} =
V_{H^+tb}^{(2)}V_{t\tilde{t}_{\alpha}\tilde{\chi}^0_j}^{(1)*}V_{b\tilde{t}_{\alpha}\tilde{\chi}^+_i}^{(2)},~~~
F_h^{(2)}=
V_{H^+tb}^{(1)}V_{t\tilde{t}_{\alpha}\tilde{\chi}^0_j}^{(2)*}V_{b\tilde{t}_{\alpha}\tilde{\chi}^+_i}^{(1)}.
\end{eqnarray}
Then the form factors contributed by Fig.1(c2) can be obtained from the
Eq.(\ref{c1}) by doing the following exchanges:
\begin{eqnarray}
\Lambda^{L,R}_{(2)} = \Lambda^{L,R}_{(1)}(F^{(1)}\rightarrow F^{(2)}, m_t\leftrightarrow m_b,
m_{\tilde{t}_{\alpha}} \leftrightarrow m_{\tilde{b}_{\alpha}}).
\end{eqnarray}
\item For diagram Fig.1(c3), we denote
\begin{eqnarray}
 F_a^{(3)} =V_{H^+\tilde{t}_{\alpha}\tilde{b}_{\beta}}V_{b\tilde{b}_{\beta}\tilde{\chi}^0_j}^{(2)*}
V_{b\tilde{t}_{\alpha}\tilde{\chi}^+_i}^{(1)},~~~F_b^{(3)}=V_{H^+\tilde{t}_{\alpha}\tilde{b}_{\beta}}
V_{b\tilde{b}_{\beta}\tilde{\chi}^0_j}^{(1)*} V_{b\tilde{t}_{\alpha}\tilde{\chi}^+_i}^{(2)}, \nb\\
F_c^{(3)}
=V_{H^+\tilde{t}_{\alpha}\tilde{b}_{\beta}}V_{b\tilde{b}_{\beta}\tilde{\chi}^0_j}^{(1)*}
V_{b\tilde{t}_{\alpha}\tilde{\chi}^+_i}^{(1)},~~~F_d^{(3)}=V_{H^+\tilde{t}_{\alpha}\tilde{b}_{\beta}}
V_{b\tilde{b}_{\beta}\tilde{\chi}^0_j}^{(2)*}
V_{b\tilde{t}_{\alpha}\tilde{\chi}^+_i}^{(2)}.
\end{eqnarray}
The form factors from Fig.1(c3) are expressed as
\begin{eqnarray}\label{c3}
\Lambda^{L}_{(3)}&=& \frac{iN_c}{16\pi^2}\sum_{\alpha,\beta=1}^{2} \left[F_a^{(3)} m_b C_0 -
F_d^{(3)} \mch C_{11}+ (F_d^{(3)}\mch - F_c^{(3)}\mno)C_{12} \right], \nb \\
\Lambda^{R}_{(3)}&=& \Lambda^{L}_{(3)}(F_a^{(3)}\leftrightarrow
F_b^{(3)},F_c^{(3)}\leftrightarrow F_d^{(3)}),
\end{eqnarray}
with $C_{0,11,12} = C_{0,11,12}(k_1, -p_1, m_b, m_{\tilde{t}_{\alpha}}, m_{\tilde{b}_{\beta}}).$
\item For diagram Fig.1(c4), we define
\begin{eqnarray}
F_a^{(4)} =
V_{H^+\tilde{t}_{\beta}\tilde{b}_{\alpha}}V_{t\tilde{b}_{\alpha}\tilde{\chi}^+_i}^{(2)*}
V_{t\tilde{t}_{\beta}\tilde{\chi}^0_j}^{(1)},~~~
F_b^{(4)} =
V_{H^+\tilde{t}_{\beta}\tilde{b}_{\alpha}}V_{t\tilde{b}_{\alpha}\tilde{\chi}^+_i}^{(1)*}
V_{t\tilde{t}_{\beta}\tilde{\chi}^0_j}^{(2)}, \nb\\
F_c^{(4)} =
V_{H^+\tilde{t}_{\beta}\tilde{b}_{\alpha}}V_{t\tilde{b}_{\alpha}\tilde{\chi}^+_i}^{(2)*}
V_{t\tilde{t}_{\beta}\tilde{\chi}^0_j}^{(2)},~~~
F_d^{(4)} =
V_{H^+\tilde{t}_{\beta}\tilde{b}_{\alpha}}V_{t\tilde{b}_{\alpha}\tilde{\chi}^+_i}^{(1)*}
V_{t\tilde{t}_{\beta}\tilde{\chi}^0_j}^{(1)}.
\end{eqnarray}
Then the form factors contributed by Fig.1(c4) can be expressed as:
\begin{eqnarray}
\Lambda^{L,R}_{(4)} = \Lambda^{L,R}_{(3)}(F^{(3)}\rightarrow F^{(4)},
m_t\leftrightarrow m_b,m_{\tilde{t}} \leftrightarrow m_{\tilde{b}}).
\end{eqnarray}
\end{itemize}
\par
The definitions of the one-loop integrals appearing in the above
formulas are adopted from the references\cite{s13}. The numerical
calculation of the vector and tensor loop integral functions can
be traced back to four scalar loop integrals $A_{0}$, $B_{0}$,
$C_{0}$, $D_{0}$ as shown in \cite{passvelt}.

\vskip 10mm
\begin{flushleft} {\bf Figure Captions} \end{flushleft}

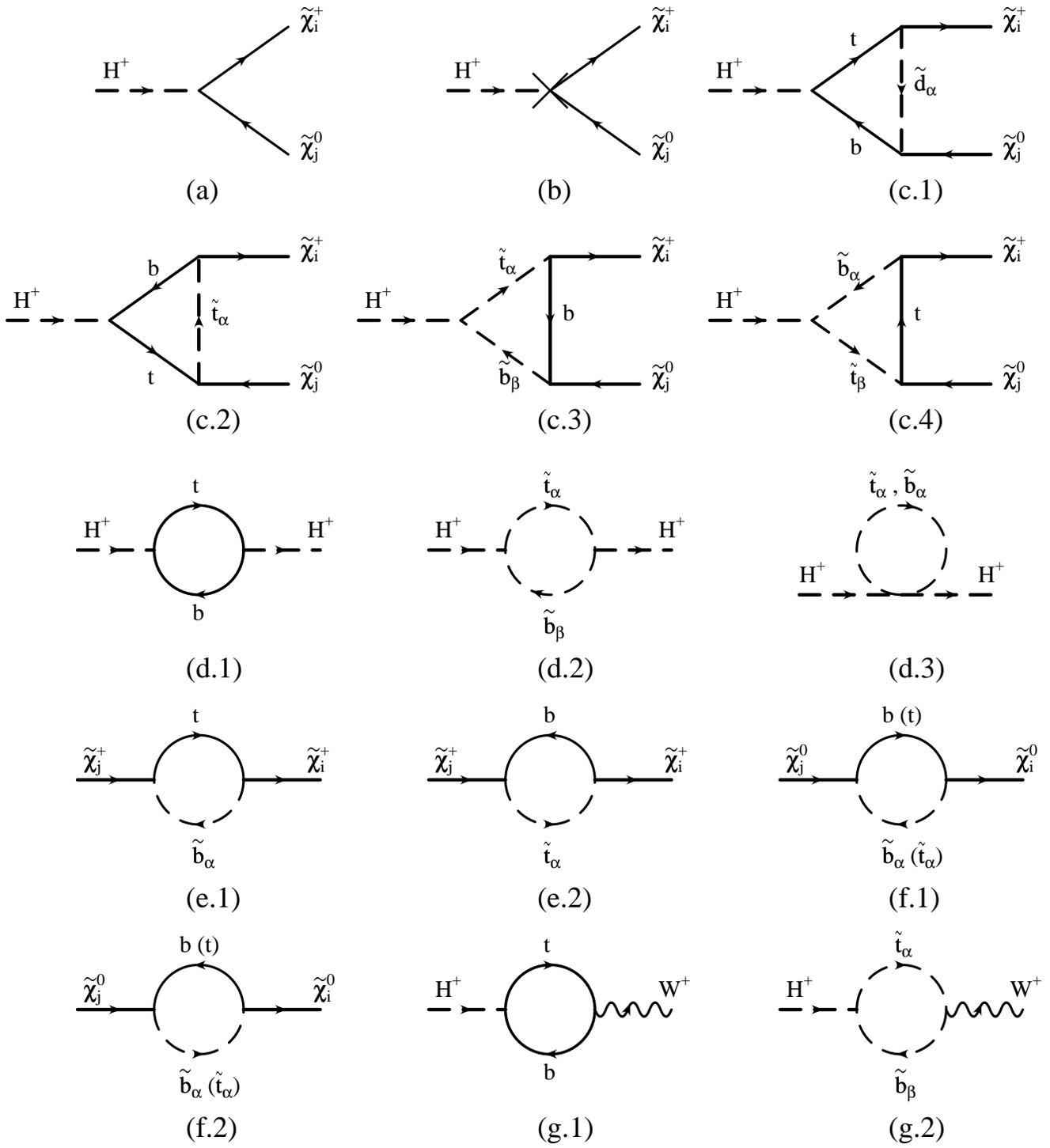
{\bf Fig.1} Feynman diagrams including one-loop quark and squark
corrections to the decay $H^+ \rightarrow
\tilde{\chi}_i^{+}\tilde{\chi}^0_j$: Fig.1(a) tree-level diagram.
Fig.1(b) counterterm for the vertex. Fig.1(c) vertex diagrams,
the lower indexes $\alpha,\beta=1,2$. Fig.1(d) charged Higgs boson
self-energies. Fig.1(e) chargino self-energies. Fig.1(f)
neutralino self-energies and Fig.1(g) are the $H^+W^+$ mixing
self-energies.
\par
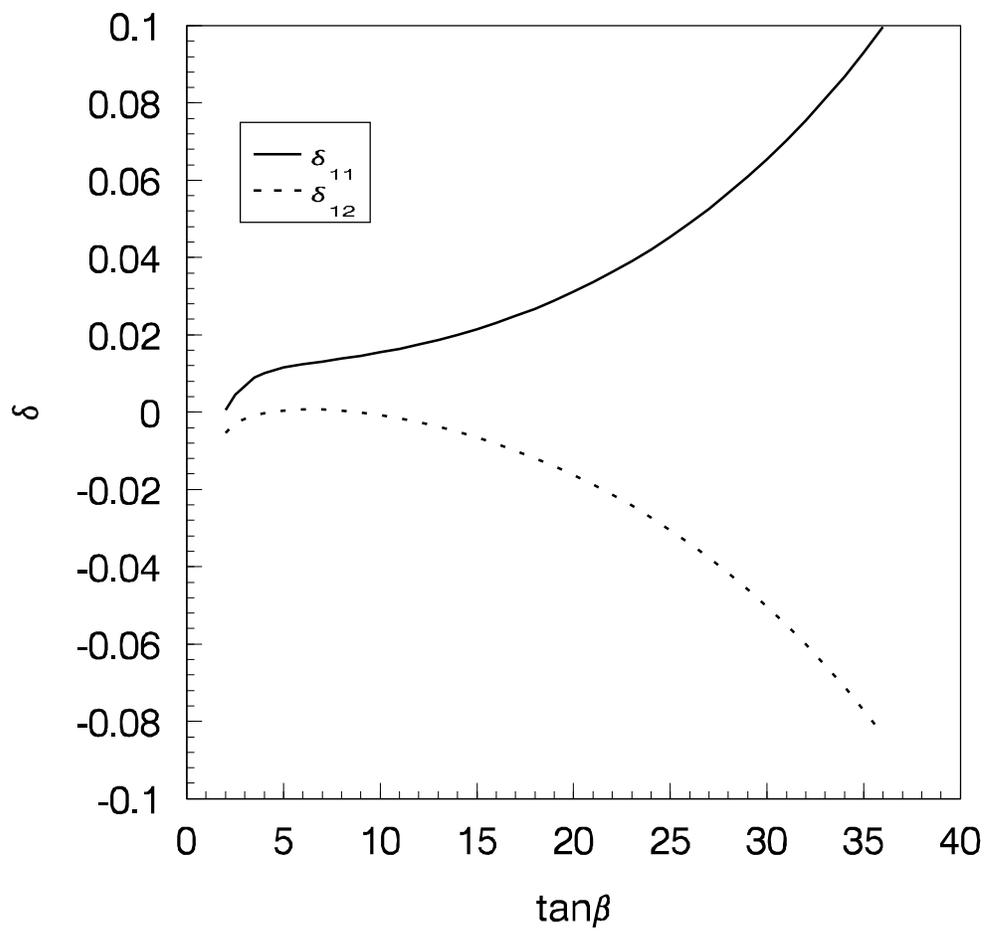
{\bf Fig.2} The relative correction as a function of $\tan\beta$
with $m_{H^+} = 250~GeV$, $m_{\tilde{\chi}^+_1} = 100~GeV$,
$m_{\tilde{\chi}^+_2} = 300~GeV$, $m_{\tilde{\chi}^0_1} = 60~GeV$,
and $m_{\tilde{Q}}=m_{\tilde{U}}= m_{\tilde{D}}=A_t=A_b=200~GeV.$
\par
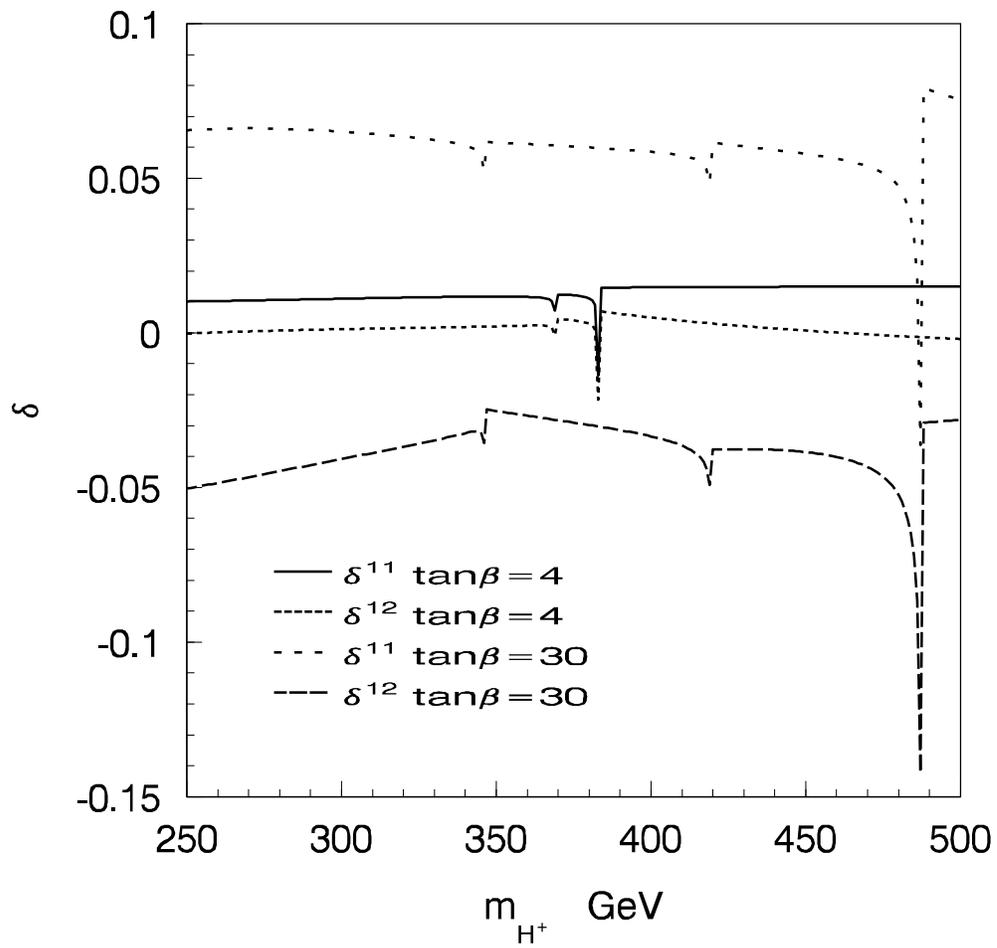
{\bf Fig.3} The relative correction as a function of charged
Higgs boson mass.
\par
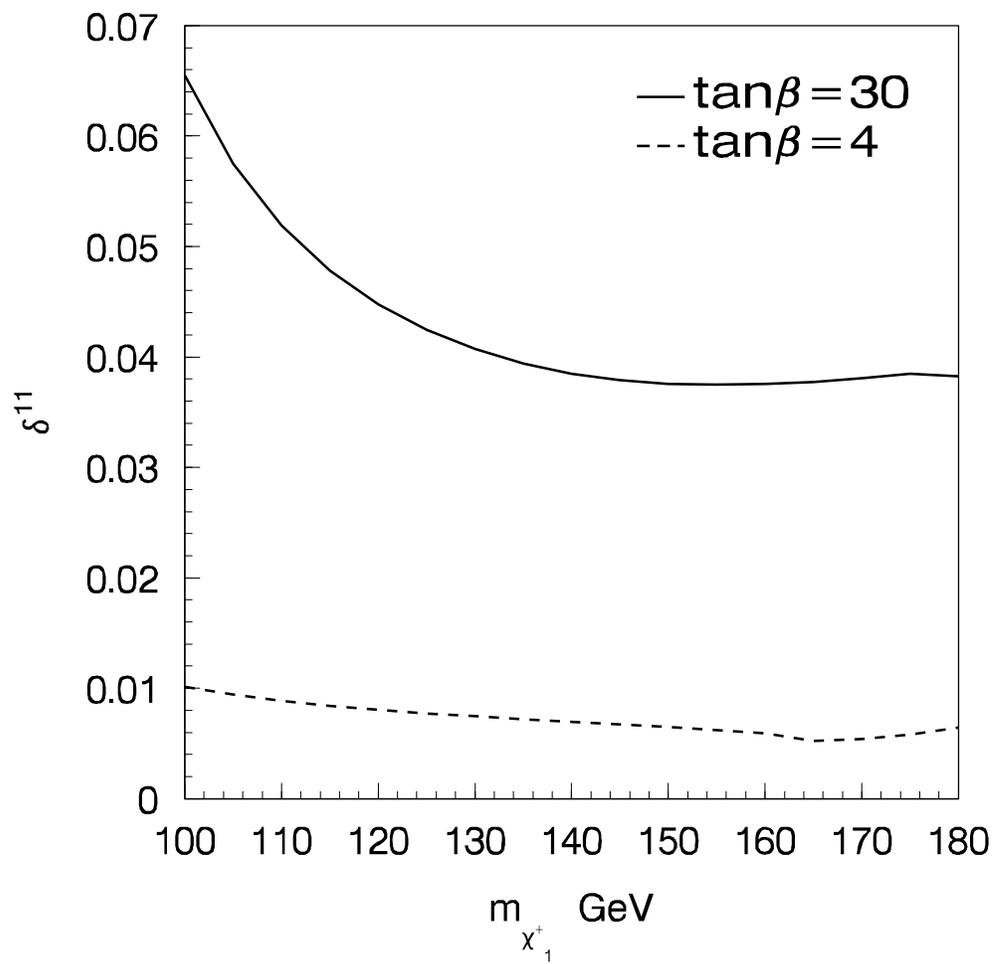
{\bf Fig.4} The relative correction $\delta^{11}$ as a function of the light
chargino mass $m_{\tilde{\chi}^{+}_1}$.
\par
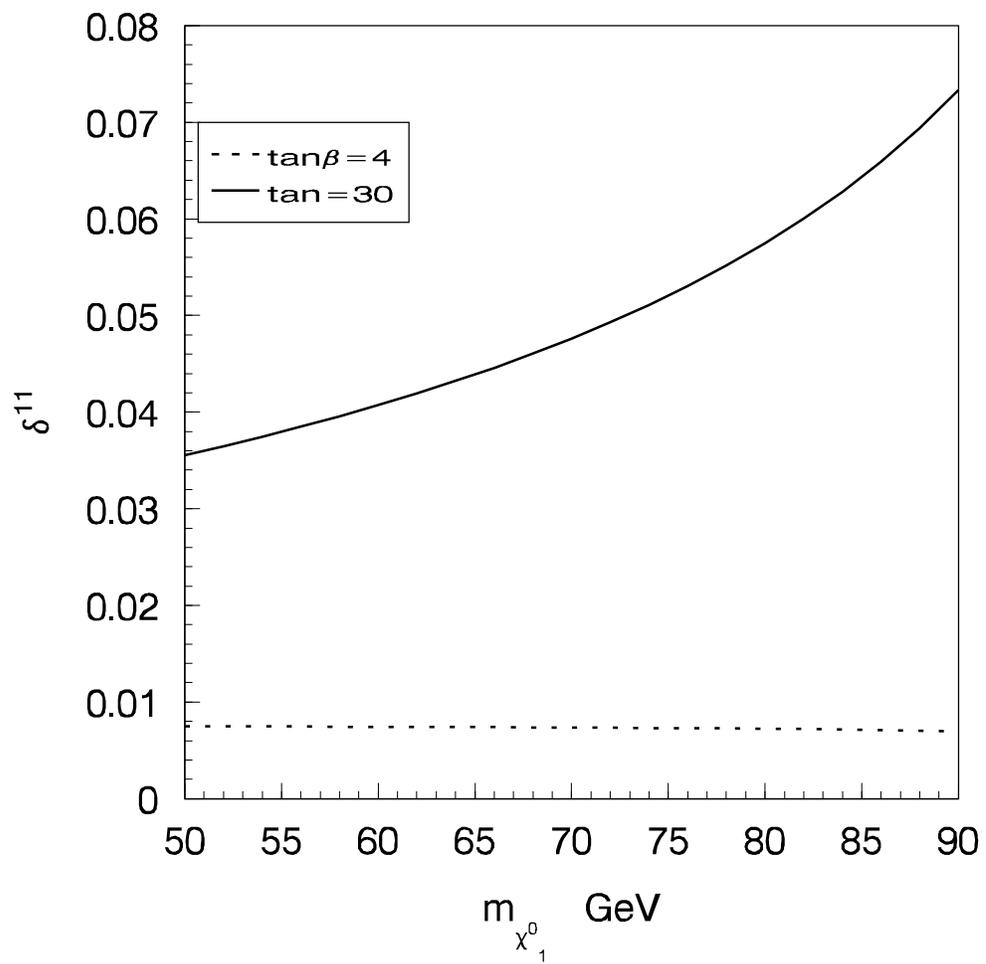
{\bf Fig.5} The relative correction $\delta^{11}$ as a function of the lightest
neutralino mass $m_{\tilde{\chi}^{0}_1}$.
\par
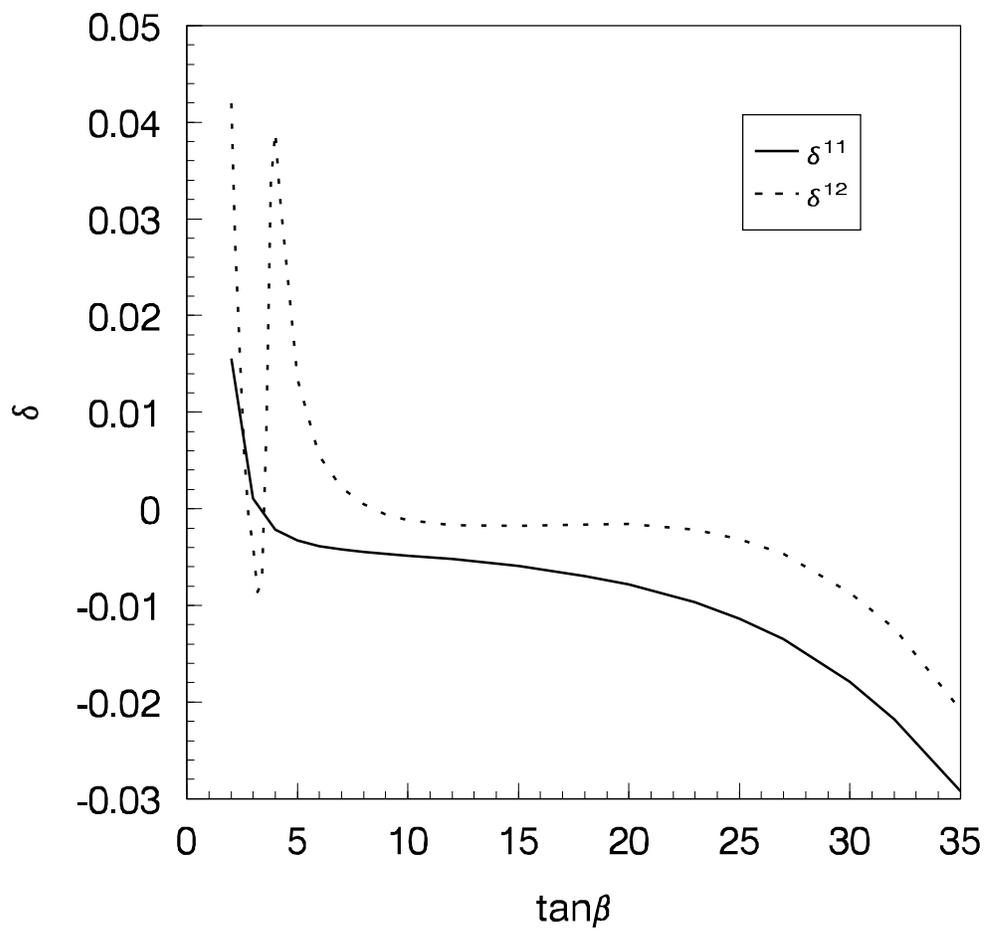
{\bf Fig.6} The relative correction as a function of $\tan\beta$ in
mSUGRA scenario.

\vskip 10mm
\begin{thebibliography}{s25}
\bibitem{s1} S.L. Glashow, Nucl. Phys. 22(1961)579; S. Weinberg, Phys. Rev.
             Lett. 1(1967)1264; A. Salam, Proc. 8th Nobel Symposium Stockholm
             1968, ed. N. Svartholm(Almquist and Wiksells, Stockholm 1968)
             p.367; H.D. Politzer, Phys. Rep. 14(1974)129.
\bibitem{s2} P.W. Higgs, Phys. Lett 12(1964)132, Phys. Rev. Lett. 13
             (1964)508; Phys.Rev. 145(1966)1156; F.Englert and R.Brout,
             Phys. Rev. Lett. 13(1964)321; G.S. Guralnik, C.R.Hagen
             and T.W.B. Kibble, Phys. Rev. Lett. 13(1964)585; T.W.B. Kibble,
             Phys. Rev. 155(1967)1554.
\bibitem{s3} H. E. Haber, G. L. Kane, Phys. Rep. 117(1985) 75.
\bibitem{eech1}  S. Komamiya, Phys. Rev. {\bf D38} (1988) 2158; A. Brignole
                et al.in Proceedings of the Workshop on $e^{+}e^{-}$ Collisions at
                $50$ GeV The Physics Potential, ed. P.M. Zerwas, DESY 92-123; A.Djouadi,
                J. Kalinowski, P.M. Zerwas, {\sl ibid} and Z. Phys. C57 (1993) 569;
                A. Arhrib, M. Capdequi Peyran\`{e}re and G. Moultaka, Phys. Lett.
                {\bf B341} (1995) 313; Marco A. Diaz and Tonnis A. ter Veldhuis
                hep-ph/9501315, DPF94 Proceedings; J. Guasch, W. Hollik and A. Kraft
                KA-TP-19-1999, hep-ph/9911452; A. Arhrib, and G. Moultaka, Nucl. Phys.
                {\bf B558} (1999) 3, hep-ph/9808317.
\bibitem{Ma1}  D. Bowser-Chao, K. Cheung, and S. Thomas, Phys. Lett.{\bf B315}, 399 (1993);
               W.G. Ma, C.S. Li ans L. Han, Phys. Rev. {\bf D53}, 1304 (1996). S.H. Zhu,
               C.S. Li and C.S. Gao, Phys. Rev. {\bf D58} (1998)055007.
\bibitem{zf1}  S. Kanemura, Eur. Phys. J. {\bf C17} (2000)473; A. Arhrib, M. C. Peyranere,
               W. Hollik and G. Moultaka, Nucl. Phys. {\bf B581} (2000)34; F. Zhou, W.G. Ma,
               Y. Jiang X.Q. Li and L.H. Wan, hep-ph/0106103.
\bibitem{Desh}N. G. Deshpande, X. Tata and D. A. Dicus, Phys. Rev. {\bf D29}
             1527 (1984); E. Eichten, I. Hinchliffe, K. Lane and C. Quigg,
             Phys. Mod. Rev. {\bf 56}, 579 (1984); {\bf 58}, 1065(E) (1986).
\bibitem{jiang}D. A. Dicus, J.L. Hewett, C. Kao and T.G. Rizzo, Phys. Rev. {\bf D40} 787(1989);
               Y. Jiang, W.G. Ma, L. Han, and Z.H. Yu, J. Phys. {\bf G24} 83 (1998);
               O. Brein and W. Hollik, Eur. Phys. J. {\bf C13} 175 (2000)
\bibitem{Gun}  J.F.Gunion, H.E. Haber, F.E. Paige, W.-K. Tung and S.S.D.
              Willenbrock, Nucl. Phys. {\bf B294} (1987) 621.
\bibitem{Diaz}  J.L. Diaz-Cruz and O.A. Sampayo, Phys. Rev. {\bf D50},
             (1994)6820.
\bibitem{Oda}  S. Moretti and K. Odagiri, Phys. Rev. {\bf D55}, (1997)5627.
\bibitem{glwh}  D.A. Dicus, J.L. Hewett, C. Kao and T.G. Rizzo, Phys. Rev.
               {\bf D40}, 789 (1989); A.A. Barrientos Bendezu and B.A. Kniehl,
               Phys. Rev. {\bf D59} (1998)015009; Phys. Rev. {\bf D61},
              (2000)097701.
\bibitem{glwh1}F. Zhou, W.G. Ma, Y. Jiang, L. Han, and L.H. Wan, Phys.
              Rev. {\bf D63} (2001)015002; O. Brein, S. Kanemura, W. Hollik,
              Phys. Rev. {\bf D63} (2001)015009.
\bibitem{glwh2}  Y.S. Yang, C.S. Li, L.G. Jin and S.H. Zhu, Phys.Rev. D62
              (2000) 095012; O. Brein, S. Kanemura, W. Hollik, hep-ph/0008308.
\bibitem{gunion2} J. Gunion, Phys. Lett. {\bf B32} 125 (1994); D.J. Miller, S. Moretti,
               D.P. Roy and W.J. Stirling Phys. Rev. {\bf D61} 055011 (2000); S. Moretti,
               Phys. Lett. {\bf B481} 49 (2000); ATLAS simulation: K.A. Assamagan and
               Y. Coadou, Preprint ATL-COM-PHYS-2000-017.
\bibitem{Assamagan} K.A. Assamagan, A. Djouadi, M. Guchait, R. Kinnunen, J.L. Kneur, D.J.
               Miller, S. Moretti, K. Odagiri and D.P. Roy, Contribution to the works
               'Physics at TeV Colliders' Les Houches, France, 8-18 June {\bf 1999},
               preprint PM/00-03, pages 36-53, February 2000, hep-ph/0002258 (to appear
               in the proceedings).
\bibitem{Bisset} M. Bisset, M. Guchait, S. Moretti, DESY-00-150, TUHEP-Th-00124,
               RAL-TR-2000-029, December 2000.
\bibitem{Copper} D.M. Copper, D.R.T. Jones and P. van Nieuwennuizen Nucl. Phys. {\bf B167}
               479(1980); W. Siegel, Phys. Lett. {\bf B84} 193(1979).
\bibitem{denner} A. Denner, Fortschr. Phys. {\bf 41} (1993) 307.
\bibitem{hollik} M. B\"{o}hm, H. Spiesberger, W. Hollik, Fortsch. Phys. {\bf 34} (1986)
               687; W. Hollik, Fortschr. Phys. {\bf 38}(1990) 165.
\bibitem{Haber2} J.F. Gunion, H.E. Haber, Nucl. Phys. {\bf B272}(1986)1.
\bibitem{Gunion3} J.F. Gunion and H.E. Haber, Nucl. Phys. {\bf B307} 445(1988);
               ibid {\bf B272} 1(1986); Errantum, ibid. {\bf B402} 567(1993).
\bibitem{Eberl} H. Eberl, M. kincel, W. Majerotto, Y. Yammada, hep-ph/0104109.
\bibitem{Denner2} A. Denner, T. Sack, Nucl. Phys. {\bf B347}, 203.
\bibitem{Kniehl} B. A. Kniehl, A. Pilaftsis, Nucl. Phys. {\bf B474} (1996) 286.
\bibitem{pdg} Particle Data Group, Eur. Phys. J. {\bf C15} 2000.
\bibitem{RGE} V. Barger, M. S. Berger and P. Ohmann, Phys. Rev. {\bf D47},
              1093(1993), {\bf D47}, 2038(1993); V. Barger, M. S. Berger,
              P. Ohmann and R. J. N. Phillips, Phys. Lett. {\bf B314},
              351(1993); V. Barger, M. S. berger and P. Ohmann, Phys. Rev.
              {\bf D49}, 4908(1994).
\bibitem{zml} M.L. Zhou, W.G. Ma, L. Han, Y. Jiang and H. Zhou, J. Phys. {\bf G25} 1641(1999)
\bibitem{s13}  Bernd A. Kniehl, Phys. Rep. {\bf 240} 211(1994).
\bibitem{passvelt}  G. Passarino and M. Veltman, Nucl. Phys. {\bf B160}, 151(1979)
\end{thebibliography}
\end{document}